\documentclass[manuscript]{aastex63}

\usepackage{array,multirow,graphicx}
\usepackage{float}
\newcolumntype{M}[1]{>{\centering\arraybackslash}m{#1}}
\graphicspath{{./}{figures/}}

\accepted{to ApJ Letters July 2020}

\shorttitle{Transmission Spectroscopy of TRAPPIST-1e}
\shortauthors{Pidhorodetska et al.}

\begin{document}

\title{Detectability of Molecular Signatures on TRAPPIST-1e through Transmission Spectroscopy 
Simulated for Future Space- Based Observatories}

\author[0000-0001-9771-7953]{Daria Pidhorodetska}
\affiliation{NASA Goddard Space Flight Center \\
8800 Greenbelt Road \\
Greenbelt, MD 20771, USA}
\affiliation{University of Maryland Baltimore County/CRESST II \\
1000 Hilltop Cir. \\
Baltimore, MD 21250, USA}
\affiliation{GSFC Sellers Exoplanet Environments Collaboration}

\author[0000-0002-5967-9631]{Thomas J. Fauchez}
\affiliation{NASA Goddard Space Flight Center \\
8800 Greenbelt Road \\
Greenbelt, MD 20771, USA}
\affiliation{Goddard Earth Sciences Technology and Research (GESTAR), Universities Space Research Association, Columbia, MD, USA}
\affiliation{GSFC Sellers Exoplanet Environments Collaboration}

\author[0000-0002-2662-5776]{Geronimo L. Villanueva}
\affiliation{NASA Goddard Space Flight Center \\
8800 Greenbelt Road \\
Greenbelt, MD 20771, USA}
\affiliation{GSFC Sellers Exoplanet Environments Collaboration}

\author[0000-0003-0354-9325]{Shawn D. Domagal-Goldman}
\affiliation{NASA Goddard Space Flight Center \\
8800 Greenbelt Road \\
Greenbelt, MD 20771, USA}
\affiliation{GSFC Sellers Exoplanet Environments Collaboration}

\author[0000-0002-5893-2471]{Ravi K. Kopparapu}
\affiliation{NASA Goddard Space Flight Center \\ 
8800 Greenbelt Road \\
Greenbelt, MD 20771, USA}
\affiliation{GSFC Sellers Exoplanet Environments Collaboration}



\begin{abstract}
Discoveries of terrestrial, Earth-sized exoplanets that lie within the habitable zone (HZ) of their host stars continue to occur at increasing rates. Transit spectroscopy can potentially enable the detection of molecular signatures from such worlds, providing an indication of the presence of an atmosphere and its chemical composition, including gases potentially indicative of a biosphere. Such planets around nearby M-dwarf stars - such as TRAPPIST-1 - provide relatively good signal, high signal/noise, and frequent transits for follow-up spectroscopy. However, even with these advantages, transit spectroscopy of terrestrial planets in the HZ of nearby M-stars will still be a challenge. Herein, we examine the potential for future space observatories to conduct such observations, using a Global Climate Model (GCM), a photochemical model, and a radiative transfer suite to simulate modern-Earth-like atmospheric boundary conditions on TRAPPIST-1e. The detectability of biosignatures on such an atmosphere via transmission spectroscopy is modeled for various instruments of $JWST$, $LUVOIR$, $HabEx$, and $Origins$. We show that only CO$_2$ at $4.3\ \mu$m would be detectable at the $> 5\ \sigma$ level in transmission spectroscopy, when clouds are included in our simulations. This is because the impacts of clouds on scale height strongly limits the detectability of molecules in the atmosphere. Synergies between space- and ground-based spectroscopy may be essential in order to overcome these difficulties.

\end{abstract}


\section{Introduction}
\label{sec:intro}
The search for small, Earth-like rocky exoplanets has made significant progress since the launch of $Kepler$ in 2009. To date, there are over 4,000 confirmed exoplanets orbiting a multitude of stars, with additional candidates and planets identified frequently. As we continue to detect terrestrial exoplanets that resemble Earth in size, some of the most exciting discoveries point towards planets that fall within the habitable zone (HZ), the region around a star where surface water oceans may be stable under the correct conditions \citep{Kasting1993,Kopparapu2013}. Such detections have thus far been biased to low mass stars (late-K and M-dwarf stars), as their small size and compact HZs give way to planets with short orbital periods that provide a high frequency of transits. It has been estimated that 16$\%$ of early M-dwarf stars and 64$\%$ of mid M-dwarf stars contain terrestrial-sized planets orbiting within the habitable zone \citep{Dressing:2015,Vanderburg2020}. While both radial velocity and transit techniques have revealed the first rocky exoplanets orbiting within the HZ of their low-mass host stars \citep{Anglada2016,Gillon2016,Gillon2017}, the majority of planets discovered through the transit method, specifically, orbit extremely close to their host star \citep{Kaltenegger2012}. This bias has influenced the discovery of many rocky planets orbiting M-type stars. Fortunately, the same observational bias that has led to the detection of these worlds will also make them more amenable to follow-up characterization, including the identification of atmospheric species via transit spectroscopy \citep{Palle2018}. \\

For some exoplanets, transit spectroscopy and/or secondary eclipse measurements (primarily done from space with the $Hubble$ $Space$ $Telescope$ $(HST)$ and the $Spitzer$ $Space$ $Telescope$) have provided empirical details on their atmospheric compositions \citep[e.g.][]{Seager:2010,Sing2016}. These investigations have primarily targeted "hot Jupiters" - gas-giant planets with orbital periods of only a few days \citep[e.g.][]{sing2011, sing2013, sing2015, pont2013, gibson2013, Nikolov2015, Evans2018}. Several sub-Neptune and super-Earth type planets have also been targeted for atmospheric characterization \citep[e.g.][]{Kreidberg2014, knutson2014, Bourrier2018, wakeford2018, Benneke2019, benneke2019b}. However, as discoveries of rocky exoplanets both within and outside of the HZ continue to increase, the first attempts to put constraints on their atmospheric properties with transit spectroscopy measurements have begun \citep[e.g.][]{deWit2016,deWit2018,southworth2017, Delrez2018, Ducrot2018,Ducrot2020,diamondlowe2018, diamondlowe2020,Burdanov2019}. \\

Within this assortment of planets, one of the most exciting and nearby exoplanetary systems that is a target for future observations is the TRAPPIST-1 system \citep{Gillon2016, Gillon2017}. Bearing seven Earth-sized exoplanets \citep{Gillon2017} orbiting an ultra-cool late-type M-dwarf star (M8V; \cite{Liebert2006}) located 12.4 parsec from Earth \citep{lindegren2018}, the TRAPPIST-1 planets are similar in size and irradiation to the rocky planets within our Solar System \citep{Gillon2017}. Their ultra-cool, low-mass parent star signifies that the evolution of their existence and the pathways they undertook to form are potentially much different than what our Solar System planets experienced \citep{Turbet2018}. This leaves us with the ideal laboratory to study how the atmospheric evolution of a planet orbiting an M-dwarf star can impact its habitability \citep{Wolf2017,Lincowski2018,Turbet2018}. Among the seven planets, 3-D climate simulations have shown that TRAPPIST-1e could be the most habitable planet of the system, able to maintain liquid water on its surface across a large range of atmospheric compositions \citep{Wolf2017,Turbet2018,Fauchez:2019a,Fauchez:2019b}. This makes it an ideal target to search for the presence of biosignatures, molecular features that may indicate evidence of life.  \\ 


 For the TRAPPIST-1 system, data obtained by $HST$ provide initial constraints on the extent and composition of the planet's atmospheres, suggesting that the four innermost planets do not have a cloud/haze-free H$_2$-dominated atmosphere \citep{deWit2016,deWit2018}. However, follow up work by \cite{Moran2018} have shown that $HST$ data can also be fit to a cloudy/hazy H$_2$-dominated atmosphere. Complementary to $HST$, NASA's $Spitzer$ $Space$ $Telescope$ - which played a major role in the discovery and orbital determination of TRAPPIST-1d, e, f, and g \citep{Gillon2017} - has also allowed us to put additional constraints on the atmospheric composition of TRAPPIST-1b. Transit observations with $Spitzer$ \citep{Delrez2018} have found a $+208\ \pm  110$ ppm difference between the 3.6 and $4.2\ \mu$m bands, suggesting CO$_2$ absorption. $Spitzer$ also showed that transit depth measurements do not show any hint of significant stellar contamination in the $4.5\ \mu$m spectral range. \cite{morris2018} reached the same conclusion using a "self-contamination" approach based on the $Spitzer$ data set. $Spitzer$'s "$Red$ $Worlds$" Program encompassed over 1000 hours of observations of the TRAPPIST-1 system, whose global results have been presented \citep{Ducrot2020}. $HST$ and $Spitzer$ measurements have also been combined with transit light curves obtained from space with $K2$ \citep{Luger2017} and from the ground with SPECULOOS-South Observatory \citep[\textit{SSO};][]{burdanov2018,gillon2018article} and Liverpool Telescope \citep[\textit{LT};][]{steele2004} where \cite{Ducrot2018} produced featureless transmission spectra for the planets in the $0.8-4.5\ \mu$m wavelength range, showing an absence of significant temporal variations of the transit depths in the visible. Additional ground-based observations with the $United$ $Kingdom$ $Infra-Red$ $Telescope$ ($UKIRT$), $Anglo-Australian$ $Telescope$ ($AAT$), and $Very$ $Large$ $Telescope$ ($VLT$) also show no substantial temporal variations of transit depths for TRAPPIST-1 b, c, e, and g \citep{Burdanov2019}. While the $K2$ optical data set detected a 3.3 day periodic 1 \% photometric modulation, it is not present in the $Spitzer$ observations \citep{Delrez2018}. Further constraints on the molecular weight and presence/absence of atmospheres on the TRAPPIST-1 planets will require additional observations with future facilities. \\ 
 
The next generation of observatories will allow for far more in-depth explorations of atmospheric properties of the TRAPPIST-1 planets. In particular, data from the $James$ $Webb$ $Space$ $Telescope$ $(JWST)$ could provide strong constraints on atmospheric temperatures and on the abundances of molecules with large absorption bands \citep{Gillon2016}.  $JWST$ houses two science instruments capable of using transit spectroscopy to detect light from planets and their host stars: The Near-Infrared Spectrograph \citep[\text{NIRSpec};][]{bagnasco2007,ferruit2014} and Mid-Infrared Instrument \citep[\text{MIRI};][]{bouchet2015} low resolution spectrometer \citep[\text{LRS};][]{kendrew2015}. NIRSpec, which will cover the infrared wavelength range from 0.6-$5.3 \ \mu$m, intends to analyze the spectrum of over 100 objects observed simultaneously. MIRI has both a camera and a spectrograph that perform between the range of 5-$28\ \mu$m, with transit observations ending at $12\ \mu$m. Only the low resolution spectroscopy (LRS) mode allows for time series observations with MIRI.  \\ 

Prior studies have evaluated the potential of $JWST$ to characterize the TRAPPIST-1 planets \citep{Lincowski2018,Lustig_Yaeger2019}. \cite{Morley2017} determined that less than 20 transits are needed to rule out a flat line with a confidence level of $5\sigma$ if the atmosphere is CO$_2$-dominated on six of the seven TRAPPIST-1 planets. However, $JWST's$ ability to characterize \textit{individual} molecular features using transit spectroscopy will be much more limited. This is partially due to the effects of clouds \citep{Suissa:2019a,Komacek:2019,Fauchez:2019a}.  CO$_2$ could be the only gas in a HZ terrestrial planet's atmosphere that can be detected \citep{Lustig_Yaeger2019,Fauchez:2019a}. For other atmospheric states, other gases may also be detectable. For example, at higher abundances, CH$_4$ \citep{Lustig_Yaeger2019}  and O$_2$-O$_2$ collision-induced absorption \citep[\text{CIA};][]{misra2014using,Lustig_Yaeger2019,Fauchez:2019c} might also be detectable by $JWST$. \\

Biosignature detection with $JWST$ has also been modeled explicitly for methanogen-dominated biospheres that produce high CH$_4$ concentrations. The combination of high CO$_2$ and high CH$_4$ is a potential biosignature for an Archean-like world \citep{Arney2016,krissansen2016detecting}, because the CH$_4$ fluxes required to sustain CH$_4$ in the presence of high CO$_2$ concentrations are orders of magnitude greater than fluxes that are consistent with geological activity. For planets with biospheres more similar to the one on modern-day Earth, $JWST$ may not be able to detect any biosignatures, if clouds are present in the planet's atmosphere \citep{Komacek:2019,Fauchez:2019a}. \\

 In 2016, the Astrophysics Division in NASA's Science Mission Directorate commissioned the study of four large concepts in preparation for the 2020 Astrophysics Decadal Survey. In this work, we simulate and compare with respect to $JWST$ the transit spectroscopy performances of three of those concept missions: the $Large$ $UV/Optical/Infrared$ $Surveyor$ \citep[\textit{LUVOIR};][]{team2019luvoir}, the $Habitable$ $Exoplanet$ $Observatory$ \citep[\textit{HabEx};][]{HabEx2018}, and $Origins$ (\cite{meixner2019origins}; formerly the Far-Infrared Surveyor). Any of the concepts that are prioritized in the Decadal Survey will have a proposed launch date in the 2030s.  \
 
 \paragraph{LUVOIR:} $LUVOIR$ is a concept for a large, multi-wavelength ($0.1-2.5\ \mu$m) serviceable observatory following the heritage of $HST$. $LUVOIR$'s current proposed architecture is a scalable observatory whose eventual size would fall within a range defined by two point design concepts: $LUVOIR$-A consists of an on-axis, large (15~m) segmented aperture telescope while $LUVOIR$-B consists of an off-axis, large (8~m) segmented aperture. The notional instrument for transit observations in the near-UV and near-IR is the High Definition Imager (HDI, 0.2-2.5 $\mu$m; \cite{team2019luvoir}).
 
 \paragraph{HabEx:} While $HabEx$ proposes a multitude of architectures, this work simulated the baseline architecture, which includes a 4-m monolithic, off-axis telescope concept with a wavelength range of 0.1-$1.8\ \mu m$. $HabEx$ is equipped with a suite of four proposed instruments that demonstrate various science capabilities, but the most relevant instrument for transit spectroscopy work is the HabEx Workhorse Camera (HWC, 0.2-$1.8\ \mu$m; \cite{HabEx2018}). 
 
 \paragraph{Origins:} The current design concept for $Origins$ is a 5.9~m on-axis telescope with a $Spitzer$-like structure that allows for minimal deployment while having a collecting area nearly equivalent in size to that of $JWST$ \citep{Battersby2018}. Observations with $Origins$ intend to cover a broad wavelength range (3-$600 \ \mu$m) with instruments that have an improved sensitivity compared to $JWST$, mainly due to the greatly reduced telescope temperature ($<$5 K). $Origins$ proposes multiple science instruments, but the most appropriate for conducting transmission spectroscopy measurements is the Mid-Infrared Spectrometer Camera-Transit Spectrometer (MISC-T, 2.8-20 $\mu$m; \cite{meixner2019origins}).  \\
 
The objective of this work is to cross-compare the capability of each of these future space-based missions to characterize a modern Earth-like TRAPPIST-1e (or an equivalent potentially habitable transiting M-dwarf exoplanet) via transmission spectroscopy. The paper is structured as follows: Section \ref{sec:method} discusses the method and the tools used in this study to simulate both the climate and the transmission spectra of TRAPPIST-1e. Sections~\ref{sec:results} presents the results of our simulations, identifying each gaseous signature in the spectra and their detectability with future observatories.  Discussions of our results are provided in Section \ref{sec:discussion}. Finally, conclusions and perspectives are presented in Section~\ref{sec:conclusions}. \\

\section{Models and Methods} \label{sec:method}
In the following subsections, we describe our methods for assessing the signal-to-noise ratio (S/N) of various molecular signatures for TRAPPIST-1e with numerous instruments and observational modes available to $JWST$, $LUVOIR$, $HabEx$, and $Origins$. Although there is a significant parameter space to be explored regarding the potential atmospheric scenarios for temperate terrestrial planets, this work examines the detectability of a modern Earth-like atmosphere, both with and without the presence of aerosols. Each atmospheric configuration is created with the LMD-Generic Global Climate Model \citep[\text{GCM};][]{Wordsworth2011}, and is then analyzed using the Planetary Spectrum Generator (PSG;  \url{https://psg.gsfc.nasa.gov}), an online radiative-transfer suite that computes synthetic transit spectra for a wide range of objects such as planets, moons, comets, and asteroids \citep{Villanueva2018}. We aim to determine how feasible the detection and characterization of a modern Earth-like atmosphere would be for $JWST$, and we assess how the feasibility of its detection may be improved with observations from future observatories. We consider an atmosphere to be detected when sufficient S/N is achieved on the strongest molecular feature in the spectrum at a 5$\sigma$ confidence level.\\

\subsection{Climate simulations with the LMD-Generic Global Climate Model (GCM)}
This work employs the 3-D Global Climate Model (GCM) developed and maintained at the Laboratoire de Météorologie Dynamique (LMD), specifically their generic GCM, LMD-G. We use LMD-G to simulate an atmosphere with the same composition as modern Earth. We use modern Earth, because it is the only example of a globally habitable planet, and is therefore also the most well-studied example of one. As such, modern Earth is the most widespread benchmark for habitable planets in the literature \citep[\text{e.g.},][]{Barstow2016,Morley2017, Lincowski2018}. \

Details on LMD-G can be found in \cite{Turbet2018,Fauchez:2019a}. In this work, we have performed climate simulations of TRAPPIST-1e using the stellar and planetary parameters from \citep{Gillon2017, Grimm2018}. Herein, TRAPPIST-1e is assumed to be fully covered by a 100~m deep ocean (aqua-planet) with a thermal inertia of $12000~J\cdot m^{-2}\cdot K^{-1}\cdot s^{-2}$ without ocean heat transport (OHT). TRAPPIST-1e is also assumed to be in synchronous rotation. \ 
The  horizontal resolution of the model is  $64\times 48$ coordinates in longitude $\times$ latitude (e.g., $5.6^\circ \times 3.8^\circ$). In the vertical direction, the atmosphere is discretized in 26 distinct layers using the hybrid $\sigma$ coordinates (with the top of the model at $10^{-5}$ bar) while the ocean is discretized in 18 layers. The stellar TRAPPIST-1 emission spectrum was computed using the synthetic BT-Settl spectrum \citep{Rajpurohit2013} assuming a temperature of 2500~K, a surface gravity of $10^3\ m\cdot s^{-2}$, and a metallicity of 0 dex. These stellar parameters have been selected to be consistent with the \cite{Turbet2017a} and \cite{Fauchez:2019a} simulations of the TRAPPIST-1 system with the LMD-G GCM. \ 

\subsection{Photochemistry simulations with the Atmos Model}
\begin{figure}[H]
    \centering
    \resizebox{15cm}{!}{\includegraphics{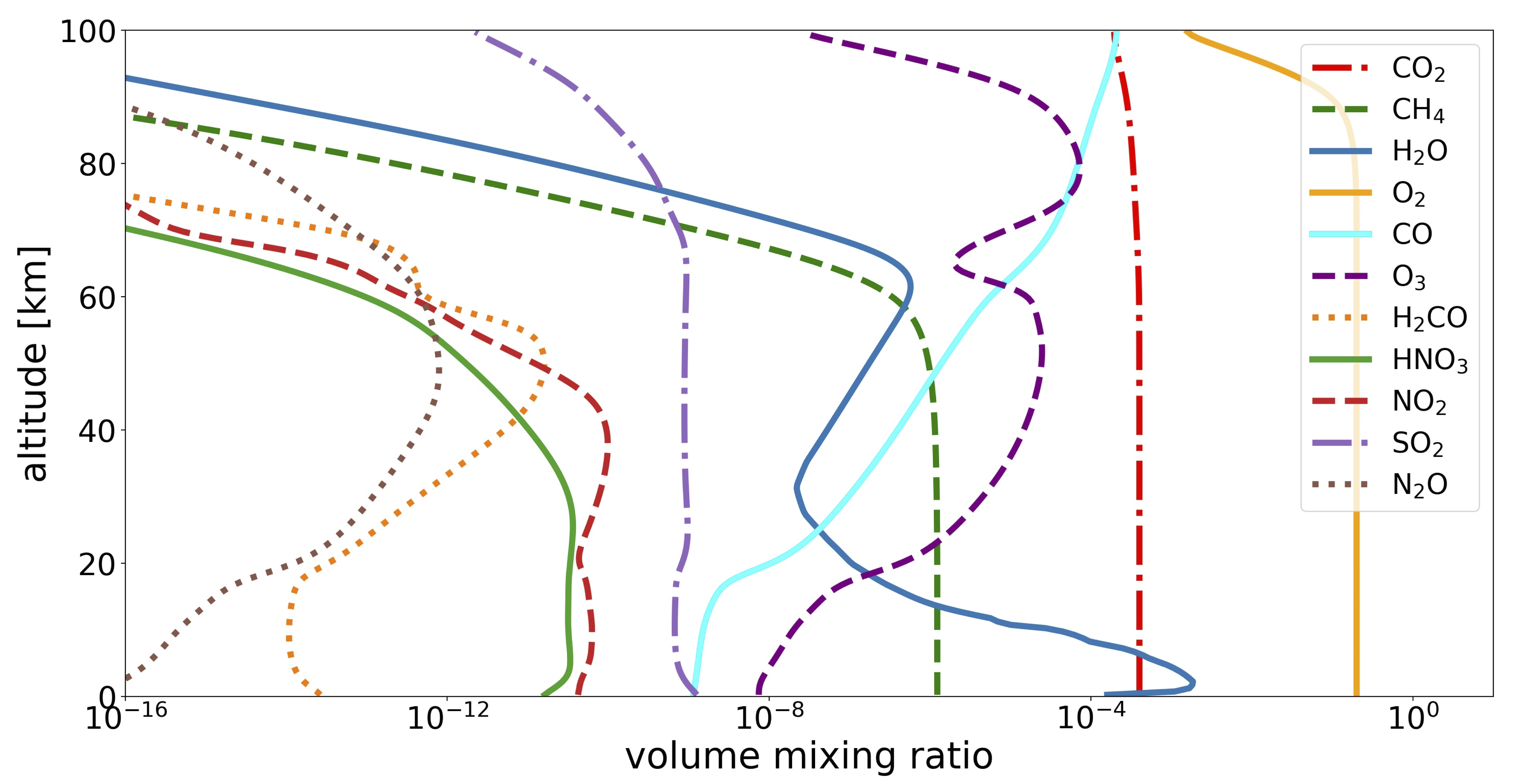}}
    \caption{Terminator-averaged gaseous atmospheric profiles for modern Earth-like boundary conditions on TRAPPIST-1e with abiotic fluxes (except for O$_2$ with a fixed mixing ratio of 0.21) produced by Atmos. The CO$_2$ vertical mixing ratio is constant, while most other gases are depleted due to photodissociation at higher altitudes. N$_2$ is excluded here as it has a fixed mixing ratio of 0.78 and is not involved in any photochemical reactions.}
\label{fig:gasprofile}
\end{figure}

The LMD-G GCM, like most GCMs used in exoplanet research, does not include photochemistry prognostically. Therefore, in order to simulate an atmospheric composition more complex than the one provided by the GCM, we utilize a 1-D photochemistry code, Atmos, to produce vertical chemical profiles. Atmos is a 1-D radiative-convective climate model, coupled with a 1-D photochemistry model, capable of simulating a variety of atmospheric redox states \citep{Arney2016,Arney2017,Lincowski2018,Meadows2018}. The boundary conditions in Atmos for modern Earth-like planets are described in \cite{Fauchez:2019a} Table 2, adapted from \cite{Lincowski2018} in Table 8, except for the H$_2$O and cloud profiles, which have been directly provided from the LMD-G GCM outputs. Photochemistry calculations have been performed at the terminator only (longitude $\pm 90^\circ$) where the star light is transmitted through the atmosphere. Atmos uses the temperature/pressure profiles and mixing ratios from the LMD-G outputs for each latitude coordinate around the terminator as used in \cite{Fauchez:2019a}. Figure \ref{fig:gasprofile} shows the mixing ratios computed for various gases when the photochemical model has converged. Gaseous profiles at the terminator are then used to compute transmission spectra with PSG.

\subsection{The Planetary Spectrum Generator (PSG)}
To create simulated spectra from our model inputs, we use PSG, a publicly available tool found online at https://psg.gsfc.nasa.gov/ \citep{Villanueva2018}. PSG is a spectroscopic suite that integrates the latest radiative transfer methods and spectroscopic parameterizations while including a realistic treatment of multiple scattering in layer-by-layer pseudo-spherical geometry \citep{Villanueva2018}. PSG permits the ingestion of billions of spectral lines of over 1,000 molecular species from several spectroscopic repositories (e.g., HITRAN, JPL, CDMS, GSFC-Fluor). For this investigation, the molecular spectroscopy is based on the latest HITRAN database \citep{gordon:2017}, which is complemented by UV/optical data from the MPI database \citep{keller-rudek:2013}. For moderate spectral resolutions ($\lambda/\Delta\lambda <$ 5000) as those presented here, PSG applies the correlated-k technique for the radiative transfer portion, while multiple scattering from aerosols is performed by PSG using the discrete ordinates method, in which the radiation field is approximated by a discrete number of streams distributed in an angle with respect to the plane-parallel normal. \\

PSG allows the user to explore many different instrument modes across a multitude of observatories. Using pre-loaded templates, PSG accounts for system throughput and presents the user with a table that describes these values, all of which are found in the detector parameters. In all synthetic spectra shown within this work, the resolving power (R) selected for each instrument is varied to accommodate the most efficient detection of the strongest molecular feature found within the spectrum of a given atmosphere. Each simulation is then plotted over a higher resolution (R = 100) model. We compute the visibility of TRAPPIST-1e within a 5-year timespan, giving us 85 observable transits. This value is fixed for each S/N calculation. 
\\

To estimate the S/N of a specific spectral line, we subtract the relative transit depth value of the line peak ($\delta$$\Delta$$_l$) from the transit depth value of the nearest continuum ($\delta$$\Delta$$_c$) - the latter could significantly vary in the VIS due to the Rayleigh slope - giving us our signal (S). We then divide S by the value of the noise of the relative transit depth (N${^{\delta\Delta}}_l$) of the line peak (l). We mathematically express this as: 

\begin{equation}
\mathrm{S} / \mathrm{N}=\left(\delta \Delta_{l}-\delta \Delta_{\mathrm{c}}\right) / {\mathrm{N}^{\delta
\Delta}}_l
\end{equation}

In order to properly capture the diversity of atmospheric conditions at the terminator as computed by the GCM, the transit spectra presented in this work were computed by running PSG at each lat-lon bin at the terminator of the planet. Information about temperature, pressure and abundance profiles at each lat-lon gridpoint from the GCM were ported into the input parameters for the spectroscopic simulations performed with PSG. These individual transit spectra were then averaged to compute the total planetary transit spectra. Considering that the spacing of the latitudinal points is constant in the GCM, the integration weights for each spectrum were assumed to be  equal, and a simple average of the transit spectra was performed.

\section{Results}\label{sec:results}
\subsection{Identification of Spectral Lines for a Modern Earth-like Atmosphere on TRAPPIST-1e}

\begin{figure}[H]
    \centering
    \resizebox{15cm}{!}{\includegraphics{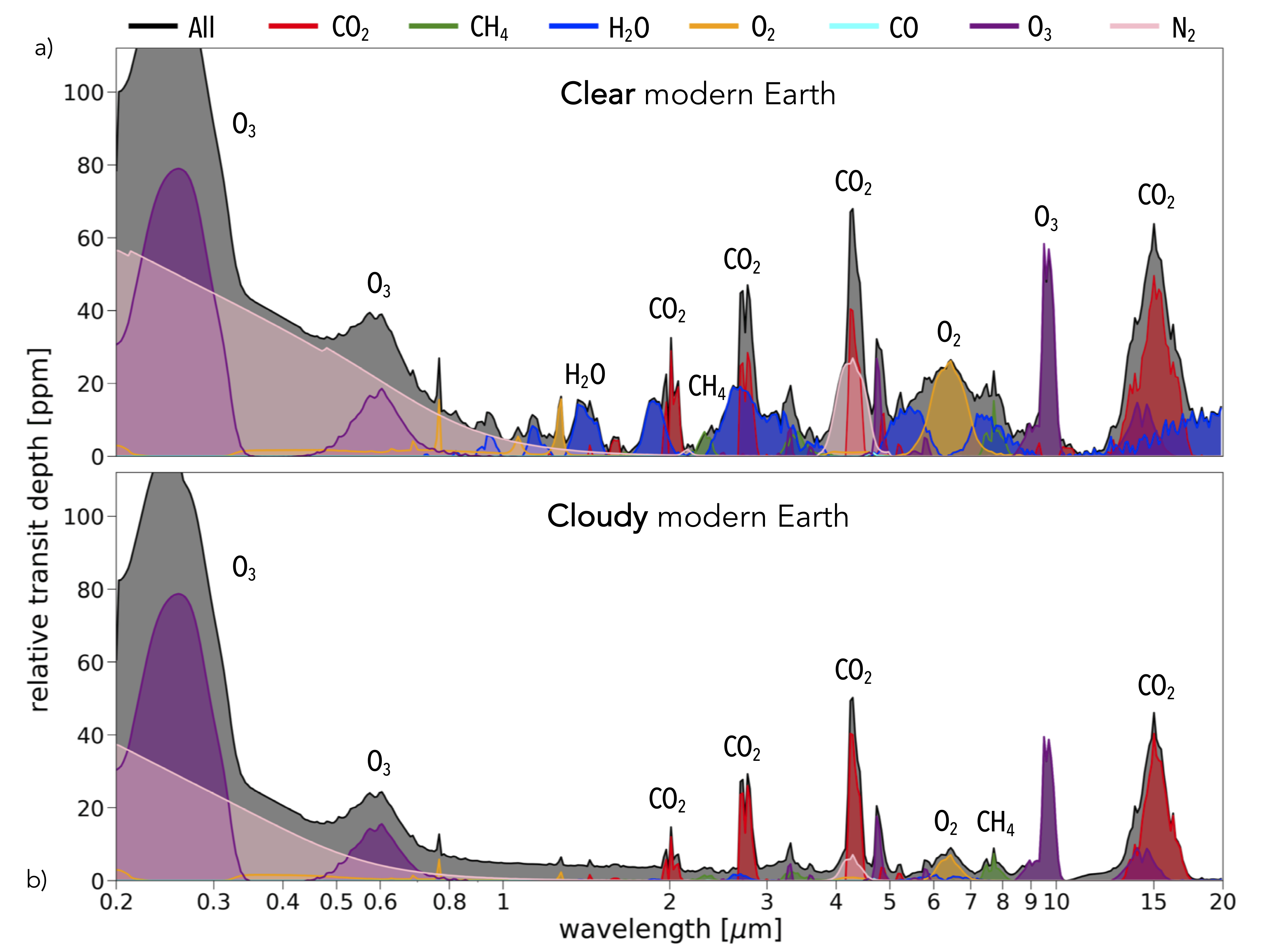}}
    \caption{Simulated transmission spectrum of a modern Earth-like atmosphere on TRAPPIST-1e both in a clear-sky scenario (panel A) and in the presence of clouds placed by the LMD-G GCM (panel B). The sum of all gases is represented by a grey shade and the individual gaseous absorption of molecular features are represented by their corresponding colors.}
\label{fig:profiles_spectra_fig_1}
\end{figure}

Figure \ref{fig:profiles_spectra_fig_1} represents the transmission spectrum assuming a clear-sky atmosphere (panel A) and a cloudy atmosphere (panel B). Only the features corresponding to the most abundant species in the vertical gas profiles in Fig. \ref{fig:gasprofile} are  shown in panels A and B. Each molecule's set of features is expressed by a unique color while contributing to the grey area beneath the black line that corresponds to the total transmission spectrum. In the UV and visible, O$_3$ and N$_2$ (via Rayleigh scattering) are the main contributors to the spectrum. In the near and mid-infrared, many wide H$_2$O absorption bands are present, along with some weaker CH$_4$ bands. The CO$_2$ features have the strongest relative transit depth comparable to the O$_3$ feature at 9.6 $\mu$m. Note that two collision-induced absorption (CIA) features are particularly notable on the spectrum: the N$_2$-N$_2$ CIA at 4.3 $\mu$m \citep{Schwieterman2015,misra2014using} and the O$_2$-O$_2$ CIA at 6.4 $\mu$m \citep{Fauchez:2019c}. The former overlaps the strong CO$_2$ feature and will be detectable only in the absence of CO$_2$. Panel B is similar in model setup to panel A, except that panel B includes the spectral effects of clouds whose locations are predicted by the LMD-G GCM, placing them at 15 km. By comparing these effects to the clear-sky atmosphere, one can see a significant decrease in the relative transit depth of each line. It is noted that clouds are strongly opaque to the visible and infrared transmitted radiations. As a result, the spectral continuum is raised above the cloud deck where the atmosphere is semi-transparent \citep{Fauchez:2019a,Suissa:2019a,Suissa:2019b}. Because the relative transit depth corresponds to the transit depth in the continuum subtracted from the transit depth in the line, a higher continuum reduces the relative transit depth. \\ 

H$_2$O is extremely affected by the presence of clouds because the H$_2$O vapor is mostly trapped beneath the cloud deck. This can also be seen in the H$_2$O profile in Fig. \ref{fig:gasprofile}. Other gases are more well-mixed up to high altitudes far above the cloud deck and are therefore are much less impacted by clouds than H$_2$O. Clouds are expected to be a recurrent feature of the atmospheres of terrestrial planets in the HZ, as liquid water on the surface would eventually evaporate and condense into the atmosphere. The opacity of clouds in the spectra poses a major obstacle in the atmospheric characterization of such planets \citep{Komacek:2019}. \\

\subsection{Detectability of a Modern Earth-like Atmosphere with Future Space-Based Observatories}
Future space-based observatories such as $JWST$ or concepts such as $HabEx$, $LUVOIR$ and $Origins$ would require the use of transmission spectroscopy to characterize the atmosphere of planets orbiting in the HZ of ultra-cool M dwarfs such as TRAPPIST-1. Direct imaging would not be possible for such close-in systems because of their inner working angle (IWA) and temperatures that are too cold to be characterizable via direct imaging \citep{Lincowski2018,Lustig_Yaeger2019,Fauchez:2019a}. Fortunately, each of these future observatories would have at least one instrument with transmission spectroscopy capabilities. The characteristics of these instruments are summarized in Table \ref{tab:instruments}. \ 

\begin{figure}[H]
    \centering
    \resizebox{16.5cm}{!}{\includegraphics{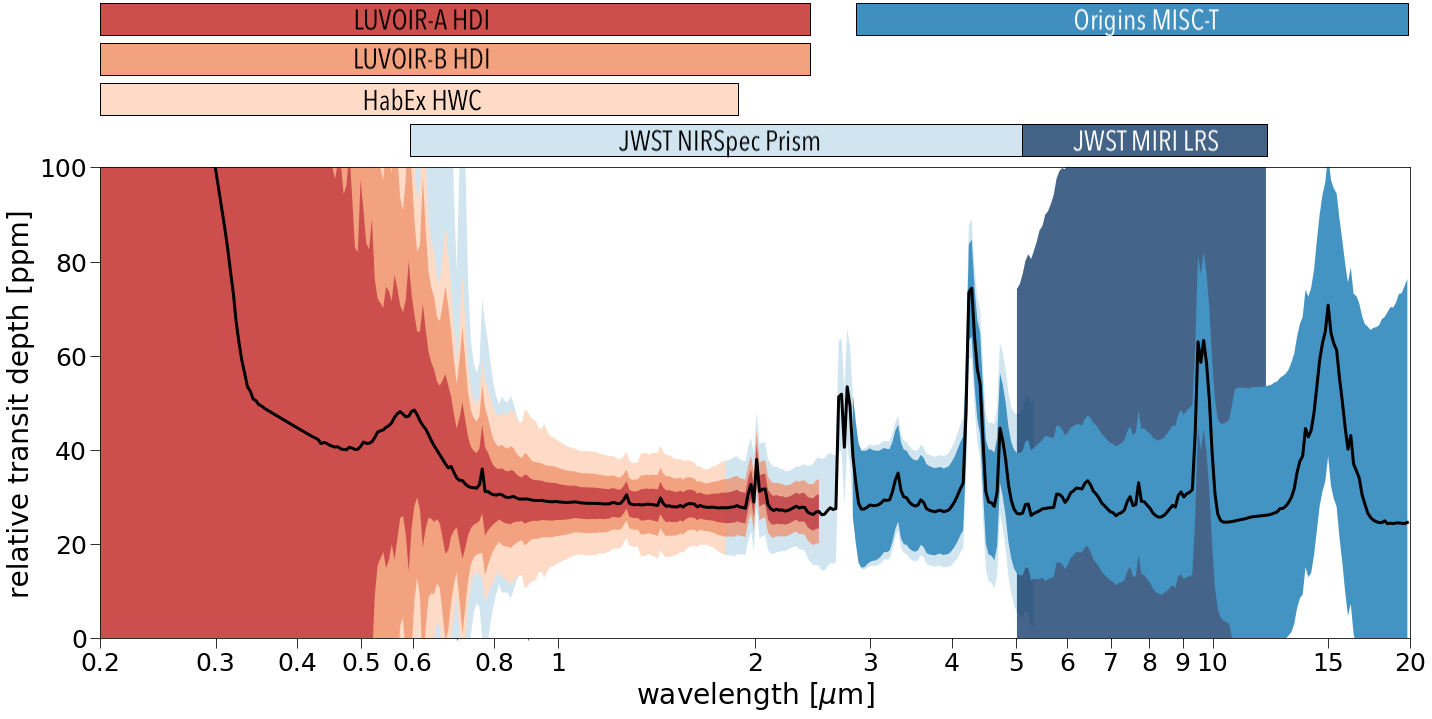}}
    \caption{Simulated transmission spectrum for a cloudy-sky, modern Earth-like atmosphere on TRAPPIST-1e as would be observed with LUVOIR-A/B HDI, HabEx HWC, JWST NIRSpec Prism/MIRI LRS, and Origins MISC-T. The relative transit depth (black line), instrumental noise expectations (spectrum colors), and instrumental wavelength ranges (colored bars) are presented. }
\label{fig:instruments}
\end{figure}

\begin{table}[ht!]
\centering
\caption{Wavelength range, instrument resolving power (R) and effective (Eff.) aperture size for $JWST$ (NIRSpec Prism and MIRI), $LUVOIR$, $HabEx$, and $Origins$}
\begin{tabular}{c c c c c c c}
\hline
\hline
Telescope & \multicolumn{2}{c}{JWST} & Origins & HabEx & LUVOIR-B & LUVOIR-A \\
\hline 
Instrument & NIRSpec Prism & MIRI LRS & MISC-T & HWC & \multicolumn{2}{c}{HDI} \\
Wavelength range ($\mu$m) & 0.6 -  5.3  & 5.0 - 12.0 & 2.8 – 20  & 0.2 - 1.8  & \multicolumn{2}{c}{0.2 - 2.5} \\
R & 100 & 100 & 50-100 & 1000 & \multicolumn{2}{c}{500 - 50,000}\\
Eff. aperture (m) & \multicolumn{2}{c}{5.6} & 5.9  & 4.0 & 8.0 & 15.0 \\
\hline
\hline
\label{tab:instruments}
\end{tabular}
\end{table}
Figure \ref{fig:instruments} shows a simulated transmission spectrum for a cloudy-sky, modern Earth-like atmosphere on TRAPPIST-1e with the addition of the wavelength range covered by each instrument, as well as their noise expectations for 85 transits assuming a photon noise limited scenario. We see that depending on the telescopes and/or instruments, different spectral lines would be detectable. For instance, while LUVOIR has the largest aperture, its wavelength coverage (cf. Table \ref{tab:instruments}) does not include the strongest CO$_2$ bands at 2.7 or 4.3 $\mu$m, and it operates in the spectral region where cloud opacity is the most prominent. As a result, the water and O$_2$ lines in this region are far too shallow to be detectable even with the largest aperture size.

When instrument performances are compared, several parameters would be at play to detect specific molecular species:

\begin{itemize}
    \item \textbf{Wavelength coverage:} Different spectral lines are accessible depending on the wavelength coverage of the instrument. 
    \item \textbf{Resolving power (R):}  $\lambda/\delta\lambda$  with $\lambda$ the wavelength and $\delta\lambda$ the spectral resolution.  Reducing R allows to increase the number of photons per spectral bands, reducing the noise. However, R should be high enough to spectrally resolve the width of the spectral feature. In this work we have optimized the resolving power by finding the lowest R to maximize the S/N.
    \item \textbf{Aperture size:} A larger aperture size collects more photons improving the S/N and therefore reducing the integration time needed to detect a given spectral feature.
    \item \textbf{Instrumental noise:} Noise produced by the instruments and the optics are wavelength dependent. Different technologies are used between $JWST$, $Origins$, $HabeX$ and $LUVOIR$-A/B that control the S/N. Such large telescopes will quickly acquire a significant number of photons after only a few transits and the noise from the source will largely dominate the total noise.
\end{itemize}
\ 

\begin{figure}[H]
    \centering
    \resizebox{19.5cm}{!}{\includegraphics{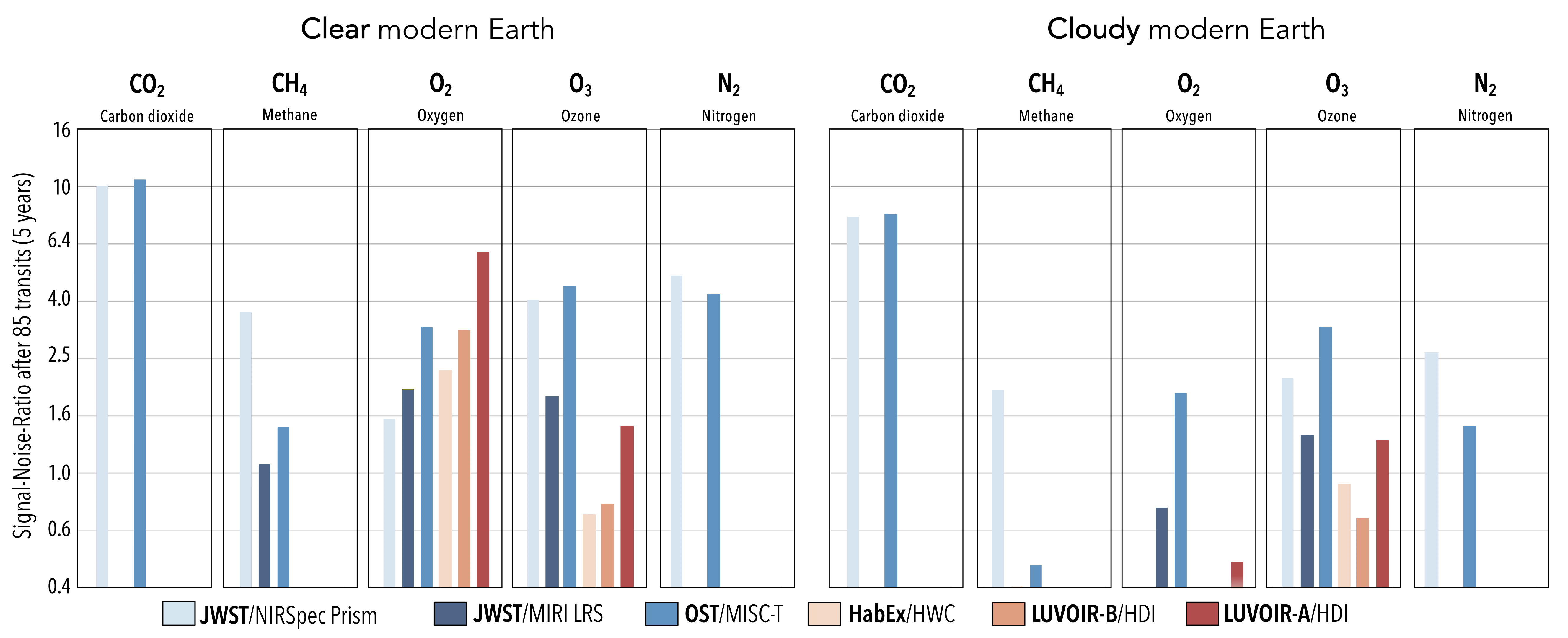}}
    \caption{Comparison of Signal-Noise-Ratios (S/N) for the different molecular indicators and observatories assuming photon noise statistics. The values were computed assuming a 5-year time span, corresponding to a total of 85 observable transits of TRAPPIST-1e.}
\label{fig:transits_instruments}
\end{figure}

Figure \ref{fig:transits_instruments} shows the signal-noise-ratios for each spectral feature after 85 transits of TRAPPIST-1e, corresponding to the expected visibility of the planet in a 5-year timespan. The clear sky atmosphere is shown for comparison but the analysis is performed on the cloudy atmosphere as it is the most realistic scenario for a habitable planet. In the presence of clouds, only CO$_2$ is detectable at a confidence level of $5\ \sigma$ or more, for $JWST$'s NIRSpec Prism and $Origins$' MISC-T. This detection is possible due to the strong 4.3 $\mu$m CO$_2$ line \citep{Lustig_Yaeger2019,Fauchez:2019a}. \\ 

\ 
For the detection of O$_3$, the S/N of $LUVOIR$-A is twice as large as that of $LUVOIR$-B due to the increase of the aperture size from 8 to 15~m. Indeed, the S/N is proportional to the square root of the number $n$ of collected photons ($\sqrt{n}$). Yet, the increase in the number $n$ of photons collected depends on the ratio between the radius squared of the two mirrors (R$_A$/R$_B$)$^2$ (assumed to be perfect disks), with R$_A$ the radius of $LUVOIR$-A (7.5~m) and R$_B$ the radius of $LUVOIR$-B (4~m). The S/N therefore improves by a factor R$_A$/R$_B$=7.5/4=1.875 between $LUVOIR$-B and $LUVOIR$-A. \\ 

Beyond collecting area and spectral coverage (photon noise statistics), the detector noise performance and systematic/calibration effects (e.g., "noise floors") will impact the detectability of a certain feature. For this investigation, our error analysis includes the detector properties as described by the observatory teams (e.g., read-noise [e-], dark current [e-/s], throughputs), while there is a large uncertainty on the actual limits or noise floors for these observatories. For $JWST$, \cite{Deming2009, Greene2016} have assumed 1$\sigma$ noise floors for NIRSpec Prism and MIRI LRS of 20 and 50 ppm, respectively. However, the systematic behavior of detectors is continuously being improved, and we consider these values to be conservative. Future facilities such as $Origins$, $LUVOIR$, and $HabEx$ expect to achieve instrumental noise floors of 5 ppm or better \citep{meixner2019origins, team2019luvoir, HabEx2018}. For this study, the TRAPPIST-1 system is bright enough to be dominated by photon noise statistics (not detector noise), and not bright enough (noise typically greater than 5-10 ppm) to be dominated by systematics (typically affecting low intrinsic noise and brighter sources), so detector properties and noise floors play a small role in establishing the detectability of the features presented here. Furthermore, there is a strong consensus that new calibration techniques and other analytical methods may lead to future reduction of these noise floors, and therefore we have not based our conclusions on these.

\section{Discussions}\label{sec:discussion}
While planets orbiting M-dwarf stars have the benefit of very frequent transits, they endure several issues regarding their characterization. First, the planets are so close to their host star that they are unable to be observed in direct imaging as they would lie within the instrument's inner working angle. In addition, planets in the habitable zone are too temperate to be characterized during a secondary eclipse from their emission spectra. Therefore, only transmission spectroscopy can be used to characterize such planets. However, as shown in this work and in previous studies \citep{Morley2017, Fauchez:2019a,Lustig_Yaeger2019,Suissa:2019a,Suissa:2019b,Komacek:2019}, atmospheric characterization through transmission spectroscopy would also be exceptionally challenging. At the mild temperatures of habitable planets, the atmospheric scale height is relatively small and the presence of clouds, inevitable if liquid water is present on the surface, strongly reduces the relative transit depth of all spectral features. These atmospheric transit depths could be on the order of or smaller than those due to stellar variability at certain wavelengths (spots, facula; \cite{Ducrot2018,zhang2018,rackham2018} and could therefore be difficult to disentangle. Also, the host star is so dim that the number of photons transmitted through the planet's atmosphere is orders of magnitudes lower than for planets orbiting G-dwarfs. As a result, the S/N improves slowly with additional transits. \\ 

According to our simulations, if TRAPPIST-1e has an atmospheric composition similar to modern Earth, only CO$_2$ would be detectable at 4.3 $\mu$m. CO$_2$ is only detectable by either $JWST$ or $Origins$, due to the fact that they are the only observatories with instruments capable of providing transit spectroscopy measurements that cover this wavelength range. \\ 

Synergies between instruments may be crucial in order to combine observations within various wavelength ranges and to accumulate transits over an extended period of time. For example, observations with future extremely large telescopes such as the $ELT$, $GMT$, or $TMT$ using cross-correlation techniques are promising and should be used in conjunction with transit observations from space. Although it is unlikely that ground-based observations targeting H$_2$O, CO$_2$, or CH$_4$ in Earth-like planets could compete with those taken from space \citep[e.g.][]{charbonneau2007,Kaltenegger2009}, this is not the case for molecular O$_2$ \citep{Snellen2013}. Other sources of absorption in an Earth-like atmosphere are relatively isolated from the O$_2$ bands at 0.76 $\mu$m and 1.26 $\mu$m, allowing for the possibility of detection after a few dozen transits \citep{Snellen2013}. This could potentially allow for the detection of O$_2$ on a planet such as TRAPPIST-1e.

The importance of wavelength range and resolution is demonstrated by the strength of the detections of CH$_4$. Of particular note is the stronger detections of CH$_4$ by $JWST$/NIRSpec, compared to $Origins$/MISC-T, despite the greater aperture of MISC-T. This is mainly due to the greater spectral resolution and spectral contrast achieved with $JWST$/NIRSpec. While none of these instruments or observatories can detect modern-day CH$_4$ for cloudy atmospheres, this may end up being important for planets with greater CH$_4$ concentrations, such as those thought to have been present on early Earth \citep{Arney2016,Arney2017}. This is especially important for any attempted search for biosignatures, as the combination of CH$_4$ and CO$_2$ might be the most detectable biosignature pairing for $JWST$ \citep{krissansen2018}. \\ 

\section{Conclusions and Perspectives}\label{sec:conclusions}
In this work, we have used TRAPPIST-1e, potentially the most promising target for atmospheric characterization of a planet in the HZ of a nearby M-dwarf, as a benchmark to compare expected transmission spectroscopy performances of future space-based observatories. This study does not aim to investigate the detectability of each gaseous species under various habitable conditions, such as those of Earth through time. Instead, we focus on the most well-known habitable atmospheric composition, that of modern Earth, and compare a variety of instrument capabilities to characterize individual molecular species. Our study shows that, despite the anticipation of tremendous future improvements in terms of aperture size and instrument performance, these factors would not be enough to characterize such planets via transmission spectroscopy. Indeed, most spectral lines from the gaseous species of a modern Earth-like atmosphere produce a relatively small transit depth and clouds drastically reduce their amplitude. Even for the largest aperture size of 15~m for $LUVOIR$-A, hundreds or thousands of observed transits would be required to detect molecular species at a 5 $\sigma$ confidence level via transit spectroscopy. Only CO$_2$ and its strongest feature at 4.3 $\mu$m could be detectable with a S/N $\geq 5$ in 85 transits, assuming all observable transits would be accrued during a 5 year mission. This feature would be observable with $JWST$'s NIRSpec Prism and with $Origins$' MISC-T. This spectral feature may be the only proxy available to detect the atmosphere of a rocky HZ planet through transmission spectroscopy with currently-planned or conceptualized space-based telescopes. \\

This work therefore demonstrates that transmission spectroscopy may not be an appropriate technique to characterize habitable planets with a composition similar to modern Earth around M-dwarfs with a single telescope, or even with a combination of space-based assets. Instrumental and mission-level synergies between space- and ground-based telescopes should be prioritized in order to improve our chances to characterize such planets.  \\ 

\section*{Acknowledgements}
We thank our anonymous reviewer for their thoughtful analysis of our work, as their suggestions greatly improved the strength of our manuscript. This work was performed as a part of the NASA Astrobiology Institute's Virtual Planetary Laboratory, supported by the National Aeronautics and Space Administration through the NASA Astrobiology Institute under solicitation No. 80NSSC18K0829. All authors acknowledge support from the NASA Goddard Space Flight Center (GSFC) Sellers Exoplanet Environments Collaboration (SEEC), which is funded in part by the NASA Planetary Science Division's Internal Scientist Funding Model. 

%


\software{Atmos \citep{Arney2016}, LMD-G \citep{Wordsworth2011}, PSG \citep{Villanueva2018}}

\bibliography{bibliography.bib}


\end{document}